\documentclass[]{aa}
\usepackage{graphicx}
\begin{document}
\title{Compton backscattered and primary X-rays from solar flares: angle dependent Green's function correction
for photospheric albedo}


   \author{Eduard P. Kontar \inst{1}
          \and
          Alec L. MacKinnon \inst{2}
          \and
          Richard A. Schwartz\inst{3}
          \and
          John C. Brown \inst{1}
          }

   \offprints{Eduard P. Kontar}

   \institute{Department of Physics and Astronomy, University of Glasgow,
G12 8QQ, UK \\ \email{eduard@astro.gla.ac.uk,
john@astro.gla.ac.uk}
         \and
             DACE, University of Glasgow, G3 6NH, UK\\\email{alec@astro.gla.ac.uk}
         \and
         SSAI, Science Systems \& Applications, Inc., Code 612.1, NASA/GSFC, Greenbelt, MD 20771, US\\
         \email{richard.schwartz@gsfc.nasa.gov}
             }

   \date{Received June 21, 2005; accepted September 30, 2005}

   \abstract{
The observed hard X-ray (HXR) flux spectrum $I(\epsilon)$ from
solar flares is a combination of primary bremsstrahlung photons
$I_P(\epsilon)$ with a spectrally modified component from
photospheric Compton backscatter of downward primary emission. The
latter can be significant, distorting or hiding the true features
of the primary spectrum which are key diagnostics for acceleration
and propagation of high energy electrons and of their energy
budget. For the first time in solar physics, we use a Green's
function approach to the backscatter spectral deconvolution
problem, constructing a Green's matrix including photoelectric
absorption. This approach allows spectrum-independent extraction
of the primary spectrum for several HXR flares observed by the
{\it Ramaty High Energy Solar Spectroscopic Imager} (RHESSI).  We
show that the observed and primary spectra differ very
substantially for flares with hard spectra close to the disk
centre. We show in particular that the energy dependent photon
spectral index $\gamma (\epsilon)=-d \log I/d \log \epsilon$ is
very different for $I_P(\epsilon)$ and for $I(\epsilon)$ and that
inferred mean source electron spectra ${\overline F}(E)$ differ
greatly. Even for a forward fitting of a parametric ${\overline
F}(E)$ to the data, a clear low-energy cutoff required to fit
$I(\epsilon)$ essentially disappears when the fit is to
$I_P(\epsilon)$ - i.e. when albedo correction is included. The
self-consistent correction for backscattered photons is thus shown
to be crucial in determining the energy spectra of flare
accelerated electrons, and hence their total number and energy.


   }

\titlerunning{Compton backscattered X-rays from solar flares}

   \maketitle
%

 \section{Introduction}
\label{intro} Bremsstrahlung HXR spectra of solar flares give us
direct information about the distribution of accelerated high
energy electrons propagating in dense regions of the solar
atmosphere. Though quite dense, the atmosphere above the X-ray
bremsstrahlung producing region is optically thin, but the very
dense lower photospheric layers are not and photons emitted
downwards are Compton backscattered by atomic electrons there with
an energy dependent efficiency. Photons of energy above $\sim 100$
keV penetrate so deeply that they are lost to the observer, while
at below $\sim 10$ keV they are mostly photoelectrically absorbed
by their first scattering. The resulting spectral reflectivity
thus has a broad hump in the range 10-100 keV with a maximum
around 30-40 keV. At some energies and view angles the
reflectivity approaches 100\% so the observed spectrum may have
been substantially affected by backscatter. This effect is
well-known in solar physics (and more generally in X-ray astronomy
- Magdziarz and Zdziarski 1995) and there have been several
discussions of its influence on observed X-ray spectra (Tomblin,
1969; Santangelo et al, 1973, Bai and Ramaty, 1978) , and on
electron spectra inferred from them (Johns and Lin, 1992;
Alexander and Brown, 2002), now that high resolution data are
available. Contamination of the observed X-ray spectrum by
reflected photons leads to flattening of the spectrum and
underestimation of the spectral index (Bai and Ramaty, 1978). This
can lead to substantial  underestimation of the total extrapolated
electron energy in a flare. Spectral features connected with a low
energy cut-off in the electron distribution, of crucial diagnostic
interest for determining the acceleration mechanism and the total
fast electron energy content, are hidden or substantially changed
by the backscattered component (Zhang and Huang, 2004). Moreover,
an apparent low-energy cut-off in albedo-uncorrected electron
spectrum may not be required by data from the ``true'' primary
(albedo corrected) spectrum.

Comptonisation of prescribed primary HXRs can be straightforwardly
studied using Monte-Carlo simulations (e.g. Bai and Ramaty, 1978).
This technique is ideal for obtaining the reflected, and hence the
total observed, photon spectrum if one knows or assumes the form
of the primary X-ray spectrum (Alexander and Brown, 2002).
However, the primary spectrum is unknown and unlikely to be an
exact power-law, as is usually assumed. Even with a power-law
distribution of primary electrons a number of physical processes
exist which will cause the resultant photon distribution to
deviate from pure power-law form e.g. effects of nonuniform
ionization (Brown, 1973; Kontar et al., 2003), beam return current
losses (Zharkova and Gordovskyy, 2005) and others. Results
obtained from Monte Carlo simulation for power-law primary photon
spectra cannot be straightforwardly applied to such situations.
More sophisticated parametric modelling of the primary photon
spectrum, with more theoretical uncertainties and free parameters,
would become necessary. Therefore, an  approach independent of the
primary spectrum is required.

Understanding and modelling backscatter has become even more
important with the advent of high quality X-ray spectra from the
Ramaty High Energy Solar Spectroscopic Imager ({\it RHESSI}) with
spectral resolution as high as $\simeq 1$ keV, in combination with
uncertainties as low as a few percent (for strong flares) allowing
detailed spectral analysis over a broad X-ray range (Lin et al,
2002). A correct treatment of Compton backscatter by a cold medium
has also been important for understanding spectra from cosmic
X-ray sources (reviewed e.g. by Svensson, 1996). This led to
construction (Magdziarz \& Zdziarski, 1995) of an analytic
(piecewise continuous), angle dependent, Green's function
$G(\mu,\epsilon_o,\epsilon)$ for the probability that a photon, of
initial energy $\epsilon_o$ and drawn from a distribution
isotropic in the downward hemisphere, incident on a plane-parallel
atmosphere, is re-emitted in direction $\theta$ ($\mu =
\cos\theta$) with energy $\epsilon$. This was obtained by fitting
the results of a large set of Monte Carlo calculations, and its
form is guided, particularly in various limits, by analytical
results such as those of Illarionov et al, (1979), Lightman et al.
(1981), White et al. (1988). Representing as it does the results
of Monte Carlo calculations, no assumptions of single scattering
or on the form of primary spectrum are involved. The only major
assumptions in the Green's function formulation are those already
mentioned: plane geometry of the cold matter - this is well
satisfied for the photospheric conditions and downward isotropy of
the primary photon distribution.

Combined with a treatment of absorption (Morrison and McCammon,
1983) with standard solar photosphere element abundances (Anders
and Grevesse, 1989), the Green's function formulation of Compton
backscatter provides us with a powerful tool for interpreting
flare hard X-ray spectra. Firstly, it allows us to calculate
easily, without time consuming Monte Carlo simulations, the total
observable photon spectrum for any specified primary spectrum, not
just the power-law cases treated by Bai and Ramaty (1978). More
importantly, because the Green's function may be used to express
the observed spectrum in terms of \emph{any} primary spectrum, we
may use regularized inversion techniques (Craig and Brown 1986,
Kontar et al, 2004, and references therein) to deconvolve the
primary photon spectrum from observations. Features of this
primary spectrum arise only from features of the HXR source
electron distribution, not of the energy-dependence of
backscattering, so it may be safely inverted (Kontar et al, 2004)
to yield a meaningful source-averaged electron distribution in the
sense of (Brown et al., 2003). Although the Green's function makes
such model-independent inversion possible, this step is carried
out here for the first time.

Section 2, 3 describe the Green's function formulation and the
incorporation of absorption respectively. Section~4 describes how
to recover the primary spectrum from data, and Section 5 applies
this to real RHESSI data. In Section~6 we discuss applications and
implications of our work.

\section{Backscattering of X-rays}

As discussed in Section~\ref{intro}, downward-emitted photons will
be either absorbed or scattered, and some scattered toward the
observer, adding to the total flux detected. The backscattered
photons will be emitted from an extended region of the
photosphere, out to the horizon distance but with greatest
intensity in an area of extent comparable to the altitude of the
primary source (Brown, McClymont and van Beek, 1975; Bai and
Ramaty, 1978; Schmahl and Hurford, 2004). The Green's function
implemented here ignores spatial structure, however, simply
summing over all photons that emerge in the correct direction.
This makes it ideal for dealing with spatially unresolved spectra.

While the Green's function (Magdziarz \& Zdziarski, 1995) was
obtained for use in studying cosmic X-ray sources, it applies to
any plane-stratified slab of 'cold' material (i.e. the scattering
electrons are non-relativistic). Scattering takes place only on
electrons, whether free or atomic. To account for elements heavier
than hydrogen the Compton cross-section is multiplied by an
effective mean atomic number z=1.2. The detailed density structure
of the medium is irrelevant (Tomblin, 1972).

Absorption, on the other hand, does depend strongly on chemical
composition, and the best estimate of photospheric abundances
should be included when adding absorption to the Green's function
treatment. We use standard abundances given by Anders and Grevesse
(1989). The heavy elements Fe/Ni play the most important role in
the range from $10$~keV up to $30$~keV, while lighter elements
contribute below $10$~keV (Morrison and McCammon, 1983).

\section{Green's function}

\label{greensec} The definition of the Green's function allows us
to write the secondary, backscattered spectrum $I_S$ for any
isotropic primary spectrum $I_{P}$ (both have units of photons
keV$^{-1}$ s$^{-1}$ cm$^{-2}$):
\begin{equation}\label{eq1}
I_{S}(\epsilon)=\int\limits_{\epsilon}^{\epsilon_{max}}I_{P}
G(\mu,\epsilon,\epsilon_0)d\epsilon_0
\end{equation}
where $G(\mu,\epsilon,\epsilon_0)$ is the angle-dependent Green's
function and $\mu =\cos (\theta)$, where $\theta$ is the
heliocentric position angle of the X-ray source. The observed
spectrum, at direction $\mu$, is then the sum of $I_S$ and $I_P$.
The importance of not averaging over viewing angle may be seen
from Bai and Ramaty (1978). The Green's function itself is shown
in Figure 1a - for details see Lightman and White (1988),
Magdziarz \& Zdziarski (1995), Ross and Fabian, (1993) and
references therein.

The total observed spectrum $I(\epsilon)$ is given by the sum of
$I_P$ and $I_S$, i.e. using (\ref{eq1})
\begin{equation}\label{itot}
I(\epsilon)=I_P(\epsilon)+\int\limits_{\epsilon}^{\epsilon_{max}}
I_{P}G(\mu,\epsilon,\epsilon_0)d\epsilon_0
\end{equation}
With measured $I(\epsilon)$ the primary spectrum $I_P$, which
tells us directly about the source electrons, may be obtained via
solution of the integral equation (\ref{itot}). This is a Volterra
Equation of the second kind (with $I_P$ appearing outside as well
as inside the integral, solution of which is more stable against
noise in $I$ than those of the first kind (e.g. Craig and Brown,
1986)). In addition, the kernel (Green's function) is peaked so
that the kernel function/matrix is not too far from diagonal. Thus
regularization to achieve an accurate stable solution for $I_P$
given $I$ is less troublesome than in other inverse problems such
as using $I_P(\epsilon)$ to find the source electron spectrum
following Kontar et al (2005).

\begin{figure}
\begin{center}
\includegraphics[width=80mm]{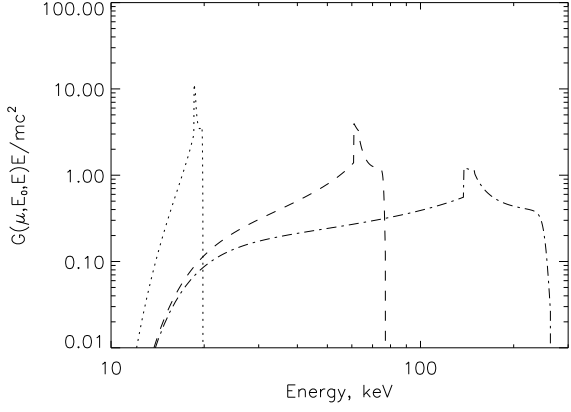}
\end{center}
\caption{Green's functions $G(\mu, \epsilon,\epsilon_0)$ including
Compton scattering and photoelectric absorption for three primary
photon energies $\epsilon _0 =20,80,300$~keV and $\mu =0.7$
($\theta \sim 45^o$).} \label{greenf}
\end{figure}

In practice measurements yield discrete quantities and the
integral equation (\ref{itot}) is presented in matrix form
\begin{equation}\label{eq2}
  I(\epsilon_i)=I_{P}(\epsilon_i)+G_{ij}(\mu)I_{P}(\epsilon_j)
\end{equation}
where we have used the summation convention for repeated indices,
and introduced what we may call the Green's matrix
\begin{equation}\label{eq3}
  G_{ij}(\mu)=\int\limits_{\epsilon_{j}}^{\epsilon_{j+1}}G(\mu,\epsilon _0,\epsilon _i)
  d\epsilon _0
\end{equation}
Note, that the integration in Eq.~(\ref{eq3}) is best performed
via a change of variable to the wavelength domain
$y=mc^2/\epsilon$, $y_0=mc^2/\epsilon _0$ due to sharp features in
the Green's function (Figure~\ref{greenf}). Special care should be
taken when integrating near $\Delta y=y-y_0=2$.

The shape of the Green's function depends on the maximum energy of
a primary photon. For primary photons with low energies $\epsilon
_{0}<30$ keV, the Green's function has a rather simple structure
close to a Dirac's delta function (Figure ~\ref{greenf}) showing
that backscattering is dominated by the first scattering, while
the contribution from higher orders of scattering is small. In the
range below $\sim 12$ keV photoelectric absorption dominates
scattering and Compton scattering can be treated as diffuse
monochromatic radiation transfer. The Green's function then can be
replaced by $G(\mu, \epsilon ,\epsilon _0)\approx
F(\mu,\lambda)\delta(\epsilon -\epsilon _0)$ (Chandrasekhar, 1960,
where the form of $F(\mu,\lambda)$ is given), $\lambda
(\epsilon)=\sigma_C(\epsilon)/(\sigma_C(\epsilon)+\sigma_A(\epsilon))$,
$\sigma_C(\epsilon)$, $\sigma_A(\epsilon)$ are the total Compton
and absorption cross-sections respectively. $\lambda$ gives the
single-scattering albedo modified by absorption and completely
controls the influence of absorption. Application to the solar
case simply requires evaluation of $\lambda$ with solar
photospheric abundances. In this energy range below 10 keV the
total Compton scattering cross-section can be approximated by the
Thompson cross-section $8\pi r_e^2/3$. The absorption
cross-section is based on the element abundances from Anders and
Grevesse (1989). We use the coefficients of a piecewise polynomial
fit to the numerical results given by Morrison and McCammon (1983)
in the range below $10$~keV and extrapolate them as
$\sigma_A(\epsilon)\sim \epsilon ^{-3}$ above $10$~keV. Due to
rapid decrease of the absorption cross-section $\sigma_A\approx
\sigma_C$ around $12$~keV. The generated Green's matrixes
(\ref{eq3}) have uncertainty around 6\% (due to the approximations
used) and $1$~keV energy resolution.

\begin{figure}
\begin{center}
\includegraphics[width=80mm]{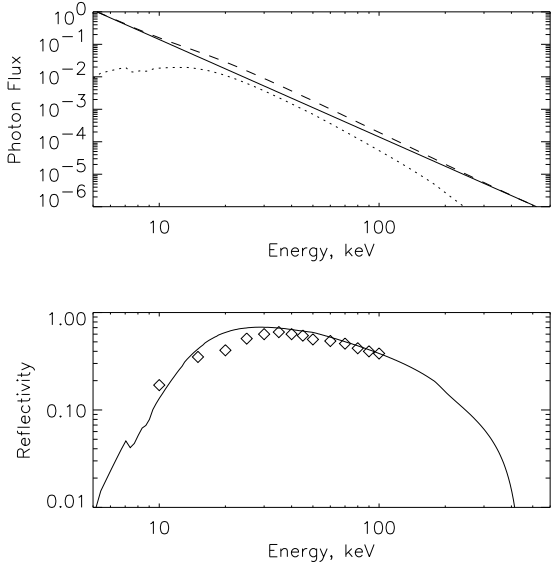}
\end{center}
\caption{Upper panel: Primary (solid line), reflected (dotted
line) and total (dashed line) photon spectra calculated assuming a
primary spectrum $I_P(\epsilon)\sim \epsilon ^{-3}$, and using the
Green's function for a X-ray source at heliocentric angle $\theta
= 45^o$. The lower panel shows the reflectivity, defined as the
ratio of reflected to primary fluxes $R(\epsilon, \theta
=45^o)=I_R(\epsilon)/I_P(\epsilon)$. The reflectivity taken from
Bai and Ramaty (1978) is shown with diamonds. Two absorption edges
of Fe at $7.1$ keV and Ni at $8.3$ keV are seen in the reflected
component.} \label{a_test}
\end{figure}

The absorption is photo-electric in nature and below 10 keV has a
complex structure with multiple absorption edges due to elements
heavier than hydrogen. In consequence it is also strongly
influenced by elemental abundances. Figure~\ref{a_test} shows the
two absorption edges of Fe at $7.1$ keV and Ni at $8.3$ keV in the
reflected component. The spectrum shows features in agreement with
the study of the reflection spectra in intergalactic nuclei by
Ross and Fabian (1993). Here we ignored the fluorescence of the
lines which is beyond the scope of our paper.

\begin{figure}
\begin{center}
\includegraphics[width=80mm]{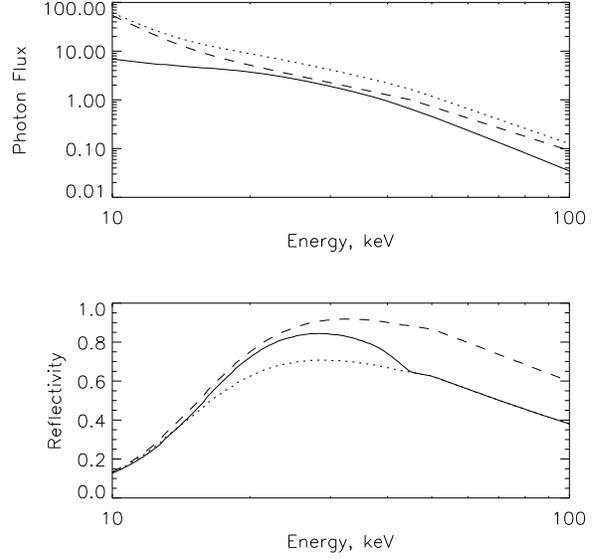}
\end{center}
\caption{Spectrum dependency of the reflected flux. Upper panel:
Primary (dashed line), reflected (solid line) and total (dotted
line) photon spectra calculated for a modelled flare spectrum
$I_P(\epsilon)\sim \epsilon ^{-3}$ for $\epsilon > 45$~keV,
$I_P(\epsilon)\sim \epsilon ^{-2}+\exp(-\epsilon/1.7)$ for
$\epsilon < 45$~keV. The lower panel shows the reflectivity for a
X-ray source at heliocentric angle $\theta = 45^o$, defined as the
ratio of reflected to primary fluxes $R(\epsilon, \theta
=45^o)=I_R(\epsilon)/I_P(\epsilon)$ (solid line). The reflectivity
assuming power-law for a primary source with spectral index 3
(dotted line) and 2 (dashed line).} \label{test_new}
\end{figure}

Note that Bai and Ramaty (1978) and Zhang and Huang (2004) assumed
a different absorption approximation from us based on earlier
photoelectric absorption cross-sections by Fireman (1974) and
photospheric abundances by Withbroe (1971). However, we should
note that the difference is rather modest accounting for around
30\% of reflectivity at low energies and becoming smaller at
higher energies, where absorption is less important (Figure
\ref{a_test}). The reflectivity, and thus the contribution of
Compton back-scattered photons into observed spectra is
spectrum-dependent (Figure \ref{test_new}). Therefore, the albedo
spectrum depends on the shape of the primary spectrum (Figure
\ref{test_new}), -- previous studies (Bai and Ramaty 1978,
Alexander and Brown, 2002) considered only the results for
prescribed power-law primary spectra. This difference is important
when one tries to infer a primary spectrum from an observed one.

We verified that we can reproduce the forward results of Bai and
Ramaty (1978) by numerically evaluating Eq.~(\ref{eq1}) for the
particular case $I_P \sim \epsilon^{-\gamma}$ and the parameters
described in Figure \ref{a_test}. Consistency with previous
treatments being established in this way, we now turn to the
inverse problem - deduction of the primary spectrum from
observations of the total spectrum.

\section{From observed to primary X-ray spectrum}
\label{invert}

In the absence of data noise we could immediately invert
Eq.~\ref{itot} to obtain $I_P$:
\begin{equation}\label{invert}
I_P(\epsilon_j)={\bf A}_{ij}I(\epsilon_i), \;\;\;\; {\bf A}=({\bf
1}+{\bf G}(\mu))^{-1}
\end{equation}
The same result as Equation (\ref{invert}) can be obtained solving
the original system of linear equations (\ref{eq3}) using, for
example, singular value decomposition. In reality, we have photon
spectra contaminated by noise. If the resulting matrix has a large
condition number (ratio of largest to smallest singular values),
then a regularized approach should be adopted (e.g. Craig and
Brown 1986, Piana et al 2002, Kontar et al, 2004). In practice, we
can incorporate the effects connected with solar albedo as a part
of instrument response.

\subsection{SPEX and albedo correction}

For a given instrument such as RHESSI, the count flux $C(\epsilon
_{i})$ is the linear composition of instrument response matrix
${\bf R}_{ij}$ and the photon flux $I(\epsilon_j)$ coming from the
Sun (Schwartz et al, 2002)
\begin{equation}\label{counts}
    C(\epsilon _{i})= {\bf R}_{ij} \;\; I(\epsilon_j)
\end{equation}
However, since we are interested in the flare primary spectra
rather than the observed flux, we can combine equations
(\ref{eq2}) and (\ref{counts}) to get
\begin{equation}\label{counts2}
    C(\epsilon _{i})= {\bf R^{\prime}}_{ij} \;\; I_p(\epsilon_j)
\end{equation}
where ${\bf R^{\prime}}_{ij}={\bf R}_{ik}({\bf 1}_{kj}+\alpha {\bf
G}_{kj})$. Here we introduced a coefficient $\alpha$ that accounts
for anisotropy of the source in the Eddington approximation - that
is, distinct but constant specific intensities in the downward and
in the upward hemispheres, in the ratio $\alpha$.

This approach has already been incorporated into the SPEX software
(Schwartz et al, 2002) including the object oriented version
(OSPEX). The software is publicly available as a part of Solar
SoftWare (SSW).

\section{RHESSI solar flare HXR spectra}
\label{data}

\subsection{Photon spectral correction}
\begin{figure}
\begin{center}
\includegraphics[width=80mm]{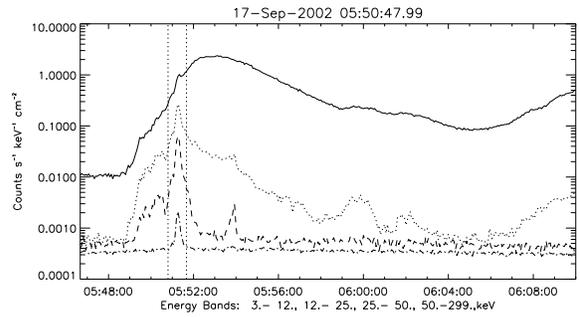}
\end{center}
\caption{Temporal variation in four energy channels (4 seconds
cadence) of the count rates summed over seven front RHESSI
detectors for the September 17, 2002 solar flare. The vertical
lines show the accumulation interval selected 5:50:48-5:51:36~UT
for further spectral analysis.} \label{flux1}
\end{figure}

The spectral effect of Compton back-scattering on HXRs is most
pronounced for harder (flatter) spectra such as in the flare of
September 17, 2002 around $05:50$UT (see Fig.~\ref{flux1}) which
we study here though quite a number of such flares have been
observed: August 20, 2002 (Kasparova et al, 2005), April 25, 2002
(Kontar and Brown, 2005). The location of our flare, which
uniquely determines the Green's function, was in the western
hemisphere at $(560,-300)$ arcseconds corresponding to a
heliocentric angle of $\theta \simeq 41^o$.

We consider the spectrum around flare maximum (5:50:48-5:51:36~UT
) when we have enough counts over the range $3$ keV up to $>100$
keV. We used 7 out of 9 front segments, excluding detectors 2 and
7 due to their low energy resolution at the time of observation
(Smith et al, 2002). The time interval is shown in Figure 4.

\begin{figure}
\begin{center}
\includegraphics[width=80mm]{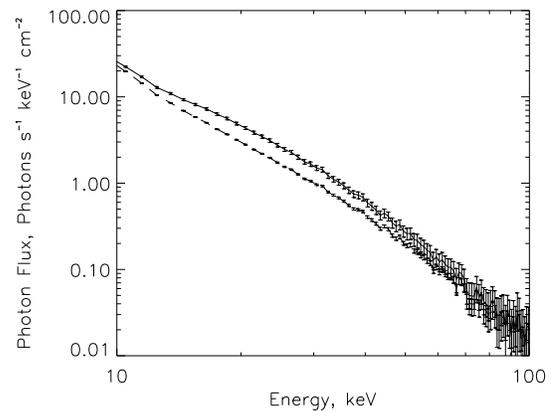}
\end{center}
\caption{Spectrum of the September 17, 2002 solar flare in the
time interval 5:50:48-5:51:36~UT. The solid line shows the
observed spectrum; the dash line is the primary spectrum
(corrected for Compton back-scattering).} \label{spectr1}
\end{figure}

After background subtraction, spectral results are shown in
Fig.~\ref{spectr1} for both the total observed spectrum, and the
primary spectrum $I_P$ determined via the method of the previous
section and using $\alpha =1$ (isotropic emission). We note how
removal of backscattered photons, as expected, reduces the primary
flux and steepens the spectrum in the 20 - 50 keV photon energy
range and slightly flattens the spectrum at energies above the
albedo maximum.

\subsection{Effect of albedo on photon spectral index}

The albedo changes the spectral indices of photon spectra, both of
power-law fits and of the local energy dependent spectral index
defined by
\begin{equation}\label{gamma}
    \gamma (\epsilon)\equiv
    - \frac{\epsilon}{I(\epsilon)}\frac{dI(\epsilon)}{d
    \epsilon}=- \frac{d \ln I(\epsilon)}{d \ln \epsilon}
\end{equation}
Direct numerical calculation of any derivative gives poor results
in the presence of data noise, even for data as good as that from
RHESSI. Differentiation is a specific case of solving an integral
equation and to obtain meaningful results in the presence of noise
needs some sort of stabilization such as Tikhonov regularization,
which we use here following Kontar and MacKinnon (2005).  We
limited our analysis to the most interesting range between $10$
and $40$ keV, where the albedo effect is most pronounced. At the
energies above $40$ keV the count rate is small and thus
uncertainties are large, whereas at the energies below $10$ keV
Fe/Ni lines affect the bremsstrahlung spectra.

\begin{figure}
\begin{center}
\includegraphics[width=80mm]{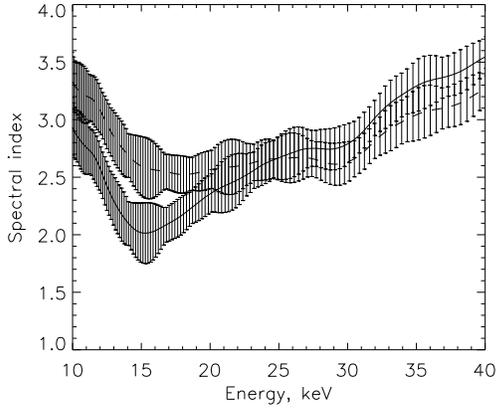}
\end{center}
\caption{Energy dependent photon spectral index $\gamma
(\epsilon)$ of the September 17, 2002 solar flare. The solid line
shows the spectral index of the observed spectrum; the dash line
is the spectral index for the primary spectrum (corrected for
Compton back-scattering). The confidence intervals represent the
range of solutions found by allowing the incident photon spectrum
to range randomly within the estimated (instrument + shot noise)
errors.} \label{gamma1}
\end{figure}

Figure (\ref{gamma1}) shows the energy dependent spectral index
for our flare spectrum. Both spectra (with and without albedo
correction) increase toward the ends of the energy interval. At
the low energies this is probably due to a thermal component while
at higher energies it is likely connected with softening of a
nonthermal electron spectrum. The spectral index of the observed
data shows a clear minimum around $15$ keV, while the spectrum
corrected for albedo has a rather extended minimum in the broad
range $15-30$~keV. The albedo correction increases the minimum
value of $\gamma $ from 2.0 to 2.6. The opposite effect can be
seen at higher energies: the spectral index at 40 keV decreases
from 3.5 to 3.2, albedo correction making the primary spectrum
harder. Qualitatively similar results have been obtained for
forward modelled spectra (Bai and Ramaty, 1978; Zhang and Huang,
2004).

\subsection{Effect of albedo on inferred source averaged electron spectra}

The primary photon flux is directly connected with the mean source
electron (so-called thin target) spectrum (Brown, 1971; Brown et
al, 2003)
\begin{equation}
\label{IFbar} I_P(\epsilon) = {1 \over 4\pi R^2} \, \, \bar n V
\int_\epsilon^\infty  \bar F(E) \, Q(\epsilon,E)\, dE, \label{def}
\end{equation}
where $Q(\epsilon, E)$ is the bremsstrahlung cross-section
differential in $\epsilon$ (Haug, 1997) and the density weighted
mean radiation source electron flux spectrum $\overline F(E)$
(electrons cm$^{-2}$ s$^{-1}$ keV$^{-1}$) is defined by
\begin{equation}
\bar F(E)= \frac{1}{\bar n V} \int_V F(E,{\bf r}) \, n({\bf r}) \,
dV.
\end{equation}
with $F(E,{\bf r})$ and $n({\bf r})$ the local electron spectrum
and source proton density at position ${\bf r}$ in radiating
volume $V$ with mean target proton density $\bar n = V^{-1}\int
n({\bf r}) \, dV$.

Solution of Equation (\ref{IFbar}) for $\overline F(E)$ is a
somewhat unstable inverse problem (Craig and Brown 1986, Piana et
al 2003) and so results are strongly affected when the observed
$I(\epsilon)$ is used instead of the albedo corrected
$I_P(\epsilon)$ required for correct solution of (\ref{IFbar}).
${\overline F}(E)$ is important in the study of acceleration
mechanisms and of the flare electron energy budget. Of particular
interest are data such as from the flare discussed above which
exhibit very flat regions which may correspond to $\overline F(E)$
with  a low energy cut-off in the mean electron spectrum which
would have major implications for electron acceleration and
propagation, and for the energy budget (Kontar and Brown, 2005).

\begin{figure}
\begin{center}
\includegraphics[width=80mm]{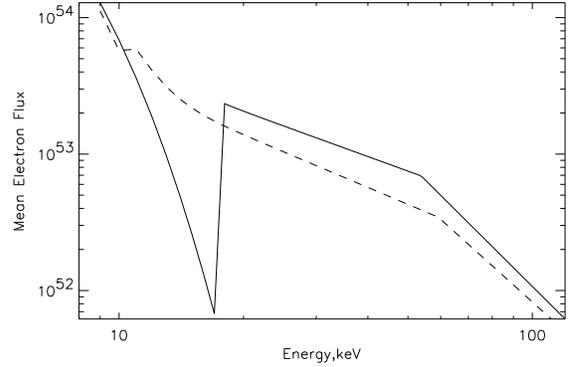}
\end{center}
\caption{Mean electron flux spectra of the September, 2002 solar
flare. The solid/dash lines show the spectrum without and with
albedo correction.} \label{spectr2}
\end{figure}

We used standard SPEX software (Schwartz et al., 2002) to fit to
the observed X-ray counts an approximate parametric $\overline
F(E)$ comprising an isothermal plus broken power-law form with low
energy cut off . The minimum $\chi ^2$ fit of the observed spectra
gives emission measure $EM=8.0\times 10^{46}$~cm$^{-3}$,
temperature $kT=1.36$~keV, spectral indexes $\delta _{low}=1.11$,
$\delta _{high}=3.01$ with break energy $E_B=54$~keV and a low
energy cut-off at $18$~keV very clearly visible above the
isothermal part (Figure \ref{spectr2}).

Next we carried out the same model fit but using the instrument
response matrix corrected for albedo as described above - i.e. we
repeat the process for $I_P$ instead of $I$. The result shows the
same tendency as we saw in the analysis of energy dependent photon
spectral index. The primary photon spectrum is softer at low
energies and harder at high energies. Now we find a slightly
bigger emission measure $EM=9.2\times 10^{46}$~cm$^{-3}$ and
slightly smaller temperature $kT=1.29$~keV. The spectral index of
energetic electrons becomes softer at low energies $\delta
_{low}=1.28$ and harder at high energies $\delta _{high}=2.73$
with the break energy at higher energies $E_B=59$keV. Most
importantly, in contrast with the fit to the observed spectrum,
the primary spectrum fit yields a much lower low energy cut-off,
around $10$ keV, not clearly visible above the large thermal
component there (Figure \ref{spectr2}). Therefore, we cannot
conclude that the true $\overline F(E)$ derived from the primary
photon spectrum has a well defined low-energy cut-off.

\begin{figure}
\begin{center}
\includegraphics[width=80mm]{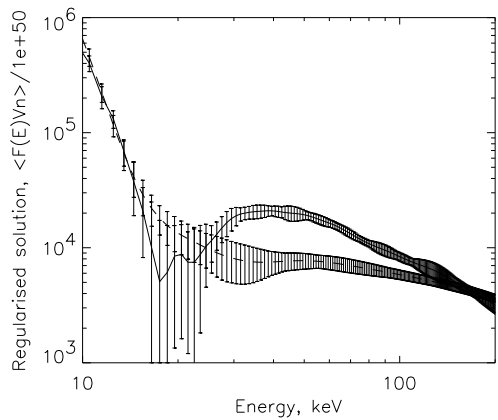}
\end{center}
\caption{Inverted mean electron flux of the August 20, 2002 solar
flare for the time interval 08:25:20-~08:25:40 UT. The dash line
shows the spectra with albedo correction. The confidence intervals
represent the range of solutions found by allowing the incident
photon spectrum to range randomly within the estimated (instrument
+ shot noise) errors.} \label{spectr3}
\end{figure}

Figure \ref{spectr3} gives a further example from the flare of
August 20, 2002 (Kasparova et al, 2005). We have obtained
regularized ${\bar F}(E)$ using the method of Kontar et al, 2005
starting from the total observed spectrum $I(\epsilon)$ and the
albedo-corrected primary spectrum $I_P(\epsilon)$. Note that a
local maximum in ${\bar F}(E)$ at $\sim 40$ keV, difficult to
accommodate in conventional thick-target models (Kontar and Brown,
2005), is no longer required when albedo is properly accounted
for.


\section{Discussion and conclusions}
\label{end}

We have shown how the Compton backscatter Green's function of
Magdziarz and Zdziarski (1995) may be employed to deduce primary
hard X-ray spectra from observations. We have applied this
procedure to RHESSI data, particularly from a couple of flares
with hard photon spectra. Without a treatment of albedo, spectral
hardening found in some flares at photon energies of $\sim 40$ keV
appears to require a local minimum in the mean fast electron
distribution (Piana et al 2003, Kasparova et al, 2005). These
local minima are particularly interesting, since they might, if
steep, (Kontar and Brown, 2005) be inconsistent with the very
widely used collision-dominated thick-target model for X-ray
production (Brown 1971; Lin and Hudson, 1976). Here we have seen
that a complete treatment of albedo removes much of the spectral
hardening in this photon energy range, potentially restoring the
viability of the collisional thick target.

The major assumption here is the isotropy of the downward directed
radiation. At the relevant photon energies here $10-100$~keV, the
intrinsic bremsstrahlung cross-section polar diagram has a
characteristic width of about $30^o$ (see figure in Massone et al,
2004). Since the emitting electron angular distribution will be
broadened by pitch-angle scattering (Leach and Petrosian, 1981;
MacKinnon and Craig, 1991), the resulting hard X-ray flux might be
fairly close to downward isotropic.

Inclusion of the albedo effect reduces the number of energetic
electrons required for the production of the observed spectra. The
total flux of energetic electrons
\begin{equation}
<vn_e n V > = \int_{E_{\rm low}}^\infty <\bar F(E)nV> \, dE,
\label{ne}
\end{equation}
where $v$ is the average electron velocity.  A thin-target fit of
the observed spectrum of the September, 2002 event gives $<vn_e n
V> = 6.2\times 10^{54}$~cm$^{-2}$sec$^{-1}$, $15$\% larger than
the value obtained for the albedo corrected spectrum . This effect
is more pronounced in the total energy than in total numbers, the
energy flux $<vEn_enV>$ being ~ 30\% overestimated when Compton
back-scattering is ignored. The influence of albedo can be even
more substantial, especially for a very flat spectra.  For example
in the August 20, 2002 flare analyzed by Kasparova et al. (2005),
the apparent low energy cut-off in observed spectra was found
around $44$~keV, a higher value than normally assumed.

It should be also stressed here that these are rather conservative
(lower limit) estimates of the albedo correction, assuming an
isotropic primary X-ray source. In fact, if the electrons are
strongly downward directed, the back-scattered photons could
produce a several times larger contribution to the observed
spectrum. This would have a major effect and clearly one must
consider the albedo to get a realistic idea of the flare electron
spectrum and energy budget. The capacity to account completely for
the effects of albedo also restores some optimism over discussing,
in terms of X-ray spectra, issues such as the lowest energies at
which electron acceleration operates (see Zhang and Huang, 2004),
though electron transport close to the thermal speed complicates
the interpretation of the X-ray spectrum there (Galloway et al.,
2005).

\begin{acknowledgements}
We are grateful to Gordon Holman and Jana Kasparova for valuable
discussions and to Kim Tolbert for adding our code to OSPEX
package. EPK, ALM, and JCB acknowledge the financial support of a
PPARC Rolling Grant. RAS is supported by contract NAS5 - 01160.
\end{acknowledgements}

\end{document}